# Weyl points enabling significant enhancement of thermoelectric performance in an antiferromagnetic van der Waals metal GdTe$_3$

Zhigang Gui[a,1], Panshuo Wang[b,1], Wenxiang Wang[a], Yuqing Zhang[a], Yanjun Li[a], Yikang Li[a], Qingyuan Liu[a], Xikai Wen[a], Qihang Liu[c], Jianjun Ying[a,d,*] and Xianhui Chen[a,d,e,*] a CAS Key Laboratory of Strongly coupled Quantum Matter Physics, and Department of Physics, University of Science and Technology of China, Hefei 230026, China.
b Key Laboratory of Material Physics, Ministry of Education, School of Physics, Zhengzhou University, Zhengzhou 450001, China.
c Department of Physics, State key laboratory of quantum functional materials, and Guangdong Basic Research Center of Excellence for Quantum Science, Southern University of Science and Technology (SUSTech), Shenzhen 518055, China.
d Hefei National Laboratory, University of Science and Technology of China, Hefei 230088, China.
e Collaborative Innovation Center of Advanced Microstructures, Nanjing University, Nanjing 210093, China.
1 These authors contributed equally to this work.
* Correspondence author: yingjj@ustc.edu.cn, chenxh@ustc.edu.cn.

The magneto-thermoelectric (MTE) effect[1] has garnered increasing attention due to the significant impact of the coupling between charge and spin on thermoelectric transport. Research on the MTE effect has demonstrated substantial improvements in thermoelectric performance, including spin fluctuations, magnon drag, and spin entropy, which indicate that magnetism plays a crucial role to enhance thermoelectric performance. This opens up new avenues for designing and optimizing thermoelectric materials, thereby breaking through the limitations of conventional thermoelectricity[2]. On the other hand, topological materials exhibit unique topological band structures that can generate anomalous thermoelectric effects through protected surface states and exotic electronic properties[3, 4]. Since the development of thermoelectric refrigeration is constrained by the scarcity of suitable solid-state cooling materials[5, 6], topological materials pave the way for breakthroughs in this field. Therefore, magnetic topological materials provide an excellent platform for investigating the MTE effect and may serve as promising candidates for solid-state cooling materials.

Recently, as a member of the rare-earth tellurides family $RTe_3$, $GdTe_3$ has attracted huge attention since it was reported as an antiferromagnetic van der Waals metal with the highest carrier mobility among layered magnetic materials, which is extremely rare. As a two-dimensional layered material, it can be readily exfoliated, making it highly suitable for applications in scientific setups involving thin films. The antiferromagnetic property provides a platform for investigating the magnetic mechanisms in thermoelectric performance and holds potential for application in spintronic devices[7]. Besides, the ultrahigh carrier mobility implies a strong response under magnetic field confirmed by large magnetoresistance[8], which also suggests the possibility of magneto response in TE performance. Previous works indicates its magnetic and topological band structures are highly tunable by the magnetic field[8]. In accordance to the exotic characteristics and practical significance, $GdTe_3$ is an ideal candidate to study MTE effect and be a promising MTE material.

In this study, we demonstrate that a magnetic field significantly enhances the thermoelectric properties of the antiferromagnetic metal $GdTe_3$. At 20 K under a magnetic field of 13.5 T, the thermopower reaches 25.2 µV K$^{-1}$, and the power factor attains 18846 µW m$^{-1}$ K$^{-2}$. These values remain unsaturated, indicating potential for even higher performance at stronger magnetic fields. Applying a 13.5 T magnetic field results in a relative enhancement of the thermopower and power factor by up to 873% and 1075%, respectively, which is nearly an order of magnitude greater than the values observed at 0 T. The absolute power factor achieved is the highest reported in metallic systems, and the magnetic field-induced enhancement is exceptionally high and rare among MTE materials. Such giant enhancement can be attributed to the magnetic field-induced topological transition from trivial metal to Weyl metal, which leads to the excellent thermoelectric performance due to the generation of Weyl points that contribute to the thermopower.

$GdTe_3$ crystallizes in a quasi-2D layered structure with an orthorhombic space group Cmcm (No. 63). It consists of double corrugated Gd-Te slabs sandwiched between two Te square-net planar sheets, as shown in Fig. 1a. The van der Waals gap exists between two neighboring Te sheets, and the stability between the Gd-Te slabs and Te sheets is maintained by Coulombic interactions due to the charge transfer between them. The synthesized plate-like single crystal of GdTe$_3$ is shown in the inset of Fig. S3 (online), with high quality and no impurity phases. Local moments of Gd become ordered around 12 K in an antiferromagnetic (AFM) state, with the magnetic moments arranged antiparallelly within the Gd-Te slab in the *ab*-plane, as illustrated in Fig. 1b. Given that the energy difference between the antiferromagnetic state and the ferromagnetic state is approximately 3 meV, the antiferromagnetic state can be readily modulated by an external magnetic field[8]. When a magnetic field is applied along the *c*-axis, the spins gradually align towards the *c* direction, forming a ferromagnetic order at high magnetic field, as shown in Fig. 1b. While the phase diagram of magnetism is intricate, featuring two antiferromagnetic transitions and a minor anomaly below 12 K, the ground state remains a robust antiferromagnetic state with antiparallel magnetic moments within the Gd-Te plane. The temperature-dependent magnetic susceptibilities are plotted in Fig. S1a (online), and the inverse magnetic susceptibility magnetization curves are also shown as the inset in Fig. S1a (online), demonstrating typical antiferromagnetic characteristics. The temperature-dependent resistivity is illustrated in Fig. S1b (online), exhibiting typical metallic behavior with a high residual resistance ratio (RRR), indicating the high quality of the single crystal. The inset of Fig. 1c shows the magnetoresistance and Hall resistivity at 2 K. Accordingly, the carrier mobility can be extracted using a two-band model (see Note S2 online), as shown in Fig. 1c. The electron and hole mobilities reach up to $10^{-1}$ m$^2$ V$^{-1}$ s$^{-1}$ at low temperature. Such high mobility is consistent with previous reports[8].

Fig. 1d-g shows the temperature dependent magneto-TE properties under a magnetic field applied along *c* axis perpendicular to temperature gradient as shown in the inset of Fig. 1d. The background is divided into two areas based on different colors that represent varying magnetic properties. The thermopower is positive without sign change implying the dominant carrier is hole. Under 0 T, the thermopower decreases monotonically. At temperatures below 100 K, the thermopower shows a relatively mild decreasing trend, whereas an abrupt reduction is observed below ~20 K. Upon application of a magnetic field, the thermopower is enhanced overall. Specifically, below 100 K, the thermopower exhibits a significant enhancement in response to the magnetic field, while the enhancement is more modest above 100 K. Thermopower reaches a maximum value of 25.2 µV K$^{-1}$ at 20 K under 13.5 T. Temperature dependent power factor (*PF*) is displayed as Fig. 1e. A modest peak forms at 20 K under 0 T owing to the alteration of decreasing tendency in *S*. Based on the square relation between *PF* and *S* (i.e., $PF = S^2\sigma$), *PF* is significantly enhanced by magnetic field showing a peak with the value of 18846 µW m$^{-1}$ K$^{-2}$. Such value is the largest *PF* value in metallic systems to our best knowledge. Contradictory to *S* and *PF*, the thermal conductivity *κ* is suppressed by magnetic field with about 20 W m$^{-1}$ K$^{-1}$ suppression around 20 K under 13.5 T, as shown in Fig. 1f, indicating the significant contribution of electron to thermal transport. Consequently, considerable improvement is achieved in *z*T with the maximum value ~0.0098 at ~ 20 K under 13.5 T.

To further elucidate the mechanism behind the enhancement of thermoelectric performance under a magnetic field, we investigate the electronic structure of GdTe₃ using first-principles calculations. In the AFM state, the spins lie within the *ab*-plane, as shown in Fig. 1b. Upon application of a magnetic field along the *c*-axis, the spins gradually align towards the *c* direction, transitioning from the collinear AFM state to a canted AFM state, and eventually to an FM state with fully polarized spins.

In the collinear AFM state, the system preserves the combined symmetry of time reversal (T) and space inversion (P), i.e., PT symmetry, guaranteeing at least two-fold degeneracy in each energy band and resulting in topologically trivial state, as shown in Fig. 2a and Fig. 2b. When an external magnetic field is applied along the *c*-axis, the magnetic moments of Gd tilt toward the *c*-direction, breaking the PT symmetry and consequently lifting the band degeneracy. Fig. 2c displays the electronic band structure when the local magnetic moments are oriented at 45° relative to the *ab*-plane, with the inset clearly demonstrating band splitting around Γ point. This phenomenon leads to the emergence of additional Weyl points and a Lifshitz transition, as illustrated in Fig. 2d, consistent with previous reports[8].

Furthermore, we investigated the relationship between the Berry curvature per unit energy interval, i.e., d$\Omega$/d$E$, and the tilt angle of Gd's local magnetic moments, as depicted in Fig. 2e and 2f. Fig. 2e presents the variation of d$\Omega$/d$E$ with chemical potential when Gd's local magnetic moments are oriented 45° relative to the *ab*-plane. The *z*-component Berry curvature reaches its maximum value at a chemical potential $\mu$ of $-0.13$ eV, d$\Omega_{xy}$/d$E$ reaching 123 Å² (eV)$^{-1}$ at this point. The continuous enhancement of d$\Omega_{xy}$/d$E$ during the field-driven rotation of magnetic moments from the *ab*-plane to the *c*-direction reveals a magnetic-field-induced topological transition and the emergence of additional Weyl points, as evidenced by the angle-dependent d$\Omega_{xy}$/d$E$ at $\mu = -0.13$ eV shown in Fig. 2f. The non-zero Berry curvature is confirmed by Shubnikov-de Haas (SdH) oscillations, which exhibit a $\pi$ Berry phase upon entering the FM state with nontrivial band topology[8].

Consequently, the extraordinary enhancement of thermoelectric performance of metallic GdTe₃ under a magnetic field can be explained in a simply way as follows: the applied magnetic field reorients the local magnetic moments of Gd, breaking PT symmetry. This symmetry breaking induces band splitting and crossing, generating Weyl points that increase the number of linearly dispersive bands. Consequently, the enhanced carrier mobility and strengthened magnetic field response due to Weyl points collectively improve the thermopower, that is $S_{\text{Weyl}} \propto d\Omega/dE$. Specifically, we developed a phenomenological theoretical model to describe this mechanism. (see Note S6 online). After a topological transition, the total thermopower ($S$) comprises two contributions: (i) the normal band component $S_{\text{normal}}$ and (ii) the Weyl points contribution part $S_{\text{Weyl}}$:

$$S = S_{\text{normal}} + S_{\text{Weyl}}.$$

Phenomenologically, $S_{\text{normal}}$ and $S_{\text{Weyl}}$ can be respectively expressed as:

$$S_{\text{normal}} \approx \alpha_0 \rho(\text{B})$$
$$S_{\text{Weyl}} \approx \alpha_{\text{Weyl}}(\text{B})\rho(\text{B}) = c\text{B}^\gamma \rho(\text{B}),$$

where $\alpha_0$ represents the normal band contribution to the thermoelectric conductivity, $\rho(B)$ denotes the magnetoresistance, $\alpha_{\text{Weyl}}$ corresponds to the Weyl-point-derived thermoelectric conductivity, and $c$ and $\gamma$ are parameters determined by band topology. As evidenced by the mobility enhancement, the Weyl-point-derived thermoelectric conductivity $\alpha_{\text{Weyl}}$ exhibits a power-law dependence on magnetic field ($\propto B^\gamma$), rendering the topological contribution $S_{\text{Weyl}}$ increasingly dominant at higher magnetic field. Fig. 2g illustrates an excellent agreement between the magnetic field dependence of thermopower predicted by our phenomenological model ($S_{\text{total}}$) and experimental measurements ($S_{\text{exp}}$) at 20 K. As the magnetic field increases, the contribution from Weyl points ($S_{\text{Weyl}}$) rises from zero, indicating that an increasing number of Weyl points contribute to the total thermopower. And we hypothesize that this phenomenological model remains applicable under high magnetic fields until the system enters the quantum limit.

Due to the contribution from band topology, the thermoelectric performance of GdTe$_3$ has been significantly improved. To clearly elucidate the effect of the magnetic field, the temperature-dependent thermoelectric parameters are plotted under 0 and 13.5 T for comparison. As shown in Fig. 2h, the thermopower increases by 22 μV K$^{-1}$ at 20 K under 13.5 T due to the magnetic field. Although the absolute increase in the thermopower may not be as pronounced as in semiconductors, it is notably high in metallic systems, especially at low temperatures. The relative enhancement, represented by $S_{13.5\text{T}}/S_{0\text{T}}$, is particularly striking, showing an 873% increase due to the magnetic field, as indicated by the stars. The unsaturated behavior of $S$ with increasing magnetic field suggests significant potential for achieving even higher values. Despite the substantial magnetoresistance, both the absolute and relative enhancements of the power factor exhibit exciting results. The absolute enhancement of PF, as shown in Fig. 2i, reaches 16970 μW m$^{-1}$ K$^{-2}$, with a corresponding relative enhancement of 1075%, exceeding one order of magnitude, which underscores the remarkable improvement attributed to the magnetic field.

Fig. 2j summarizes the thermoelectric properties of typical TE materials[3, 9-11]. The three-dimensional comparison in the 3D column plot includes the peak power factor, its corresponding absolute thermopower, and temperature. As illustrated in the figure, the power factor, which represents the power output capacity, reaches an exceptionally high value of 18846 μW m$^{-1}$ K$^{-2}$, surpassing that of nearly all other TE materials. This value is the highest recorded in metallic systems and is second only to that of the semimetal TaP[12] among all the TE materials.

It is also essential to compare these results with those of MTE materials, given the strong response of GdTe$_3$ to magnetic fields. Fig. 2k presents the peak power factors of typical MTE materials as a 2D bar graph[3, 13-18]. The material on the left, with a yellow background, indicates an enhancement in the power factor under an external magnetic field, whereas the material on the right shows a suppression effect. Indeed, the enhancement of thermoelectric performance under magnetic fields is relatively uncommon compared to the suppression effect, as magnetic fields typically reduce the degree of spin freedom[14]. As shown in Fig. 2k, the enhancement in the power factor of GdTe$_3$ under a magnetic field far exceeds that of typical MTE materials.

The significance of our work can be evaluated from two perspectives: mechanism and application. From the standpoint of topological-enhanced TE mechanisms, it is essential to discuss within the realm of topological materials. Enhanced TE performance has been observed in topological insulators and topological semimetals, originating from distinct mechanisms. For topological insulator systems, the boundary between regions with differing topological invariants results in the formation of gapless edge states. These edge states, in addition to contributing to the thermopower alongside bulk states, further enhance TE properties[19]. Examples like $Bi_2Se_3$ and $Bi_2Te_3$ exhibit excellent TE performance. In the case of topological semimetals, high TE performance arises from their intrinsic band structure—specifically, the linear dispersion bands that yield high carrier mobility, contributing to thermopower. Materials such as Weyl semimetals $Cd_3As_2$[20] serve as compelling evidence. In this work, we discovered a novel mechanism wherein a magnetic-field-induced degeneracy creates pairs of Weyl points, enhancing thermopower in metallic systems. This phenomenon shares similarities with the intrinsic characteristics of Weyl semimetals but represents an extrinsic and tunable effect. Consequently, the state formed after the field-induced topological transition can be termed a "Weyl metal state". This new mechanism significantly expands topological TE materials beyond insulators and semimetals into metallic systems.

From the perspective of applications, the focus lies on the fundamental characteristics of two-dimensional magnetic structures. Firstly, the layered structure enables easy tuning of the band structure via exfoliation. Since a magnetic field can induce band degeneracy and crossings, more crossing points can be adjusted to approach the Fermi level, forming Weyl points that contribute to thermopower through exfoliation. Secondly, the combination of magnetism and flexibility makes it feasible to develop flexible TE devices and micro spin-caloritronic setups, both in practical applications and scientific research.

In additional, several exotic phenomena have been observed in our magneto-thermoelectric studies. First, electronic and thermoelectric transport measurements (see inset configurations in Fig. S8 online and Fig. S9a online) reveal pronounced anisotropy in both resistivity and thermopower. This strong correlation between the two parameters aligns with the theoretical relationship $S \approx \alpha/\sigma$ under appropriate approximations, as detailed in Note S4 online. Second, low-temperature thermal conductivity analysis yields a calculated Lorenz number that significantly deviates from the Sommerfeld value $L_0$. This violation of the Wiedemann-Franz law suggests dominant electron-electron scattering processes, with full analysis provided in Note S5 online.

In summary, we have observed a significant enhancement in the thermoelectric performance of $GdTe_3$ under a magnetic field. The absolute enhancements in thermopower and power factor reach 22 $\mu V\ K^{-1}$ and 16970 $\mu W\ m^{-1}\ K^{-2}$, respectively, with relative enhancements of up to 873% and 1075%, representing over an order of magnitude improvement. The maximum power factor attained is 18846 $\mu W\ m^{-1}\ K^{-2}$, which is the highest value reported in metallic systems. Both thermopower and magnetoresistance exhibit strong anisotropy and unsaturated characteristics under the magnetic field, indicating a high degree of relevance between them. We have also observed electron-electron scattering-induced violations of the Wiedemann-Franz law, highlighting the critical role of carriers in both electrical and thermal transport. Theoretically calculations indicate that under the influence of a magnetic field, the topologically trivial antiferromagnetic state can be tuned to a topological Weyl metal. This transition leads to band splitting, generating additional Weyl points that contribute to enhanced thermopower, thereby achieving excellent thermoelectric performance.

Our findings demonstrate that the emergence of Weyl points can be harnessed to improve thermoelectric properties, providing a novel strategy for the development of high-performance topological materials. Additionally, we have achieved the largest enhancement in power factor within metallic systems in GdTe$_3$, showcasing significant potential for solid-state cooling and advancing the development of thermoelectric refrigeration. This mechanism of magnetic-field-induced Lifshitz transition enhancing thermoelectric performance should be applicable to other magnetic van der Waals metals.

**Conflict of interest**
The authors declare that they have no conflict of interest.

**Acknowledgments**
This work is supported by the National Key Research and Development Program of the Ministry of Science and Technology of China (2019YFA0704900 and 2022YFA1602601), the Innovation Program for Quantum Science and Technology (Grant No. 2021ZD0302800), the National Natural Science Foundation of China (grant no. 12474133 and 12494592), CAS Project for Young Scientists in Basic Research (2022YSBR-048), the Strategic Priority Research Program of Chinese Academy of Sciences (grant no. XDB25000000), Systematic Fundamental Research Program Leveraging Major Scientific and Technological Infrastructure, Chinese Academy of Sciences under contract No. JZHKYPT-2021-08, the Anhui Initiative in Quantum Information Technologies (grant no. AHY160000), Beijing National Laboratory for Condensed Matter Physics (contract no. 2023BNLCMPKF018, Shenzhen Science and Technology Program (grant no. RCJC20221008092722009 and No. 20231117091158001). Jianjun Ying was also supported by the USTC Tang Scholar.

**Author contributions**
Xianhui Chen and Jianjun Ying conceived the project and supervised the overall research. Zhigang Gui performed the thermoelectric measurements with the help of Yuqing Zhang, Yanjun Li, Yikang Li, Qingyuan Liu M. and Xikai Wen. Wenxiang Wang grew the single crystal samples. Panshuo Wang performed the DFT calculations. Zhigang Gui and Panshuo Wang analyzed the data with help with Qihang Liu. Zhigang Gui, Jianjun Ying, and Xianhui Chen wrote the manuscript with the input from all authors.

# Figure captions

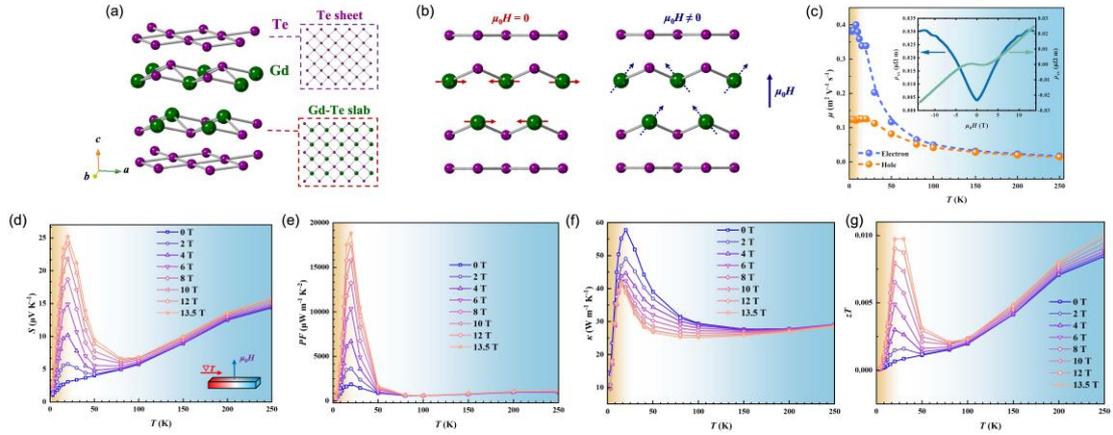

**Fig.1.** Crystal structure, magnetism, transport and magneto-thermoelectric effect of GdTe$_3$. (a) Schematic illustration of the GdTe$_3$ crystal structure with the Te square net sheet and the corrugated Gd-Te slab. (b) The magnetic structure of GdTe$_3$. When a magnetic field is applied along the $c$-axis, the in-plane magnetic moments with antiferromagnetic order gradually align towards the $c$-axis. (c) Temperature-dependent carrier mobilities. The inset shows magnetoresistance and hall resistivity at 2 K. The light yellow shading indicates the antiferromagnetic phase (AFM), while the light blue shading designates the paramagnetic phase (PM). (d) Temperature-dependent thermopower of GdTe$_3$ under different magnetic field. The inset in the bottom right corner illustrates the direction of the temperature gradient ($\nabla T$) and the magnetic field ($\mu_0 H$). (e) Temperature-dependent power factor $PF$. (f) Temperature-dependent thermal conductivity $\kappa$. (g) Temperature-dependent figure of merit $zT$ of GdTe$_3$ under different magnetic field.

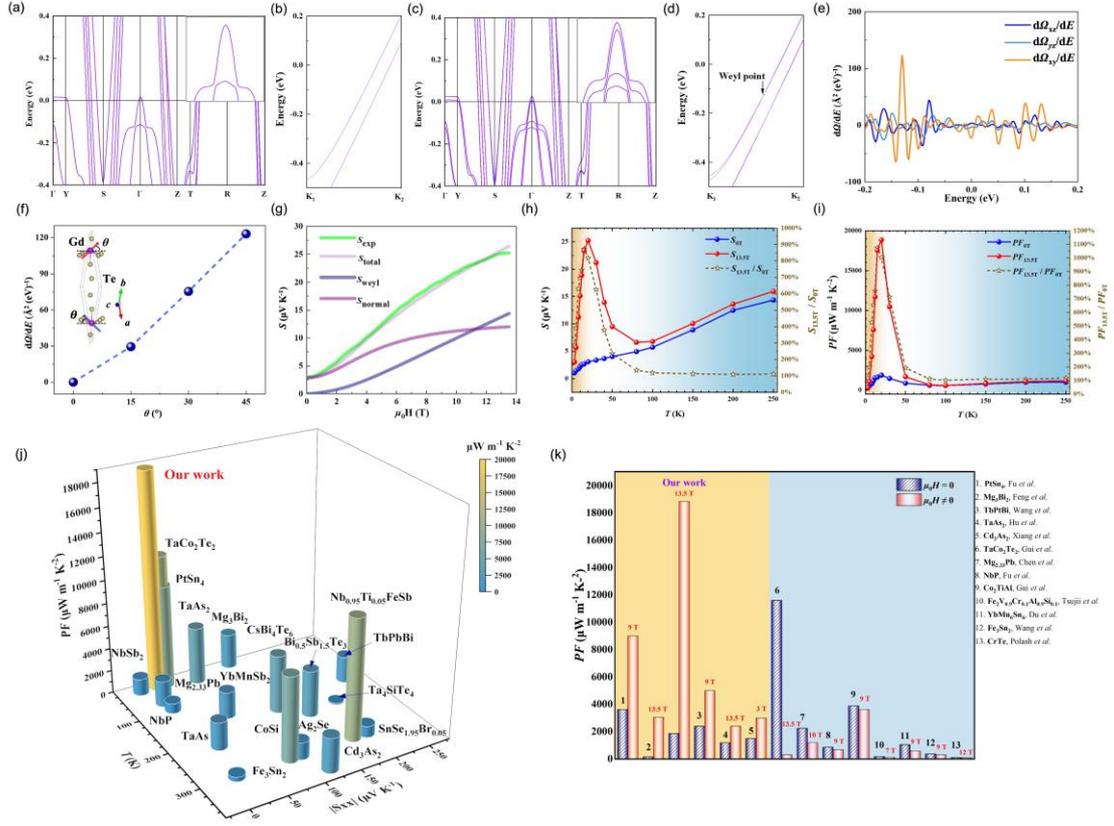

**Fig. 2.** Weyl points contribute to thermoelectric performance. The electronic band structures of GdTe$_3$ in (a) the collinear AFM state and (b) the canting AFM state. Enlarged views highlight the differences in band structures around the Γ point in these states. Panels (c) and (d) illustrate the band splitting and its evolution following a field-induced topological transition. (e) d$\Omega$/d$E$ plotted as a function of chemical potential. (f) Angle-dependent d$\Omega$/d$E$. The inset indicates the angle representing the rotation of the magnetic moment under a magnetic field applied along the $z$-direction. (g) Thermopower obtained by the phenomenological model $S_{\text{total}}$ compared with experimental results $S_{\text{exp}}$. The separated contributions of normal band component $S_{\text{normal}}$ and the Weyl points contribution part $S_{\text{Weyl}}$ are also presented. (h) Temperature-dependent thermopower under 0 and 13.5 T. The solid lines represent the absolute thermopower and the dash line indicates the ratio of $S_{13.5\text{T}}$ to $S_{0\text{T}}$. (i) Temperature-dependent power factor under 0 and 13.5 T. The solid lines represent the absolute power factor and the dash line represents the ratio of $PF_{13.5\text{T}}$ to $PF_{0\text{T}}$. (j) Comparison of peak power factor of GdTe$_3$ with other typical thermoelectric materials[3, 9-11]. (k) Comparison on magneto-thermoelectric effect among typical MTE materials[3, 13-18].